\documentclass[aps,twocolumn,superscriptaddress,nofootinbib]{revtex4}
\usepackage{graphicx}  
\usepackage{dcolumn}   
\usepackage{bm}        
\usepackage{amssymb}   
\usepackage{amsmath}
\usepackage{amsfonts}
\usepackage[colorlinks=true,linkcolor=red, citecolor=red]{hyperref}
\usepackage{tikz}
\usetikzlibrary{shapes,arrows}
\usepackage{siunitx}
\usepackage{physics}
\usepackage{hhline} 
\usepackage{booktabs}
\usepackage{subcaption}
\usepackage{float}

\usepackage{hyperref}

\begin{document}

\title{Thermodynamics and phase transitions of $\kappa$-deformed Schwarzschild-AdS black holes}

\author{A. Naveena Kumara}
\email{nathith@irb.hr}
\affiliation{Rudjer Bo\v{s}kovi\'c Institute, \\
Bijeni\v cka  c.54, HR-10000 Zagreb, Croatia}

\author{Vishnu Rajagopal}
\email{vishnu@hunnu.edu.cn}
\affiliation{Department of Physics, Key Laboratory of Low Dimensional Quantum Structures and Quantum Control of Ministry of Education, and Hunan Research Center of the Basic Discipline for Quantum Effects and Quantum Technologies, Hunan Normal University, Changsha, Hunan 410081, China} 

\author{Puxun Wu}
\email{pxwu@hunnu.edu.cn}
\affiliation{Department of Physics, Key Laboratory of Low Dimensional Quantum Structures and Quantum Control of Ministry of Education, and Hunan Research Center of the Basic Discipline for Quantum Effects and Quantum Technologies, Hunan Normal University, Changsha, Hunan 410081, China}

\begin{abstract}

We investigate the thermodynamics of the Schwarzschild-AdS black hole in the framework of \(\kappa\)-deformed non-commutative geometry by constructing an effective \(\kappa\)-deformed Schwarzschild-AdS metric from the \(\kappa\)-deformed Newtonian potential. In the extended phase space, we derive a modified first law and the corresponding Smarr relation by treating the \(\kappa\)-deformation parameter as an additional thermodynamic variable and identifying its conjugate potential. Our analysis shows that \(\kappa\)-deformation induces critical behaviour and phase transitions in an uncharged Schwarzschild-AdS black hole, with a critical ratio \(P_c v_c/T_c \simeq 0.370\) that is independent of the deformation parameter and close to the Van der Waals value. 

\end{abstract}

\maketitle

\section{Introduction}

Black holes (BHs) are among the most profound predictions of general relativity, arising as solutions of Einstein's field equations, and they provide an important arena for exploring the quantum nature of gravity. Interestingly, BHs have also been shown to behave as thermodynamic systems, with temperature and entropy directly related to the surface gravity and the area of the event horizon, respectively~\cite{Bekenstein:1973ur, Hawking:1975vcx}. In this context, BHs in asymptotically AdS space-times play a central role in understanding the foundational aspects of BH thermodynamics. This is particularly evident in \cite{Hawking:1982dh}, where the Hawking--Page phase transition between thermal radiation and a stable large BH was demonstrated for the Schwarzschild-AdS BH. This remarkable result later acquired further significance through the AdS/CFT correspondence, where it was interpreted as the confinement/deconfinement phase transition in the dual gauge theory \cite{Witten:1998zw}. Furthermore, in the extended phase space, the first law of BH thermodynamics and the corresponding Smarr relation can be consistently formulated by identifying the cosmological constant as thermodynamic pressure and its conjugate quantity as thermodynamic volume, while incorporating conserved charges such as electric charge and angular momentum. In this framework, the BH mass is interpreted as enthalpy rather than internal energy \cite{Kastor:2009wy,Kubiznak:2016qmn}. In this regard, charged AdS BHs have been shown to exhibit \(P\)-\(V\) criticality, with critical behaviour analogous to the Van der Waals phase transition in ordinary liquid--gas systems \cite{Kubiznak:2012wp}. Subsequently, critical behaviour and phase transitions of various AdS BHs in the extended phase space have been extensively investigated~\cite{Cai:2013qga,Gunasekaran:2012dq,Dehghani:2014caa,Hennigar:2015esa,Ballon-Bayona:2020xls,Singh:2020xju,Zhang:2021raw,Li:2020xkh,Cong:2021jgb,Sood:2022fio,Li:2023zfl,Astefanesei:2023sep,Yang:2024krx}. The thermodynamic framework of BH has further been enriched by interpreting geometric variables on null surfaces, deriving the physical process first law for dynamical BHs, and extending the first law to higher-curvature theories and BHs in external electromagnetic fields \cite{Chakraborty:2015wma,Chakraborty:2015hna,Mishra:2017sqs,Hu:2026slp}. 

Although Einstein gravity provides a robust description of gravitational phenomena, it is expected to break down in the strong-gravity regime, where quantum fluctuations become significant and necessitate the formulation of a quantum theory of gravity for a complete understanding of BHs. Many approaches have been developed to reconcile classical gravity with quantum theory and thereby shed light on the ultimate structure of quantum gravity. Among them, non-commutative (NC) space-times \cite{Maresca:2025sqx} have emerged as an important framework for incorporating quantum gravitational effects, in which the quantum structure of space-time is governed by an NC algebra that naturally introduces a fundamental length scale \cite{Snyder:1946qz,Doplicher:1994zv}. Recently, the NC space-time structure is shown to arise naturally from the perturbative quantum gravity, at Planck scale \cite{Frob:2022ciq,Frob:2023vay}. Over the years, various NC space-times have been proposed, and extensive studies have been carried out to incorporate NC effects into gravitational theories. Among these, Moyal space-time \cite{Douglas:2001ba} and $\kappa$-deformed space-time \cite{Arzano:2021hpg} are the two most widely studied examples.

The Moyal space-time, $[\hat{x}_{\mu}, \hat{x}_{\nu}] = i \theta_{\mu\nu}$, where $\theta_{\mu\nu}$ is a constant tensor, is a canonical NC space-time that is known to arise under the compactification of certain string theory models \cite{Seiberg:1999vs}. Most approaches based on Moyal space-time focus on deforming the Einstein--Hilbert action by replacing the ordinary product with the Moyal star product \cite{Moffat:2000gr,Aschieri:2005yw,Harikumar:2006xf}, or on deforming the standard gauge theory of gravity through the star product and the Seiberg-Witten map \cite{Chamseddine:2000si,Calmet:2005qm,Chaichian:2007dr,Juric:2025kjl}. However, NC BH solutions has also been constructed alternatively by deforming the standard BH solutions \cite{doi:10.1142/S0217751X09043353}. In \cite{Nicolini:2005vd}, NC geometry-inspired BH solutions have been constructed by replacing the standard point-like mass sources with the smeared mass distributions, resulting in the deformation of the energy-momentum tensor of Einstein's equation. Such NC-deformed BH solutions have been shown to cure the curvature singularity of standard BHs by introducing an NC-induced regular de Sitter core \cite{Nicolini:2005vd,Ansoldi:2006vg,Spallucci:2008ez,Modesto:2010rv}. Different aspects of BH thermodynamics in Moyal space-time have been extensively investigated using these deformed BH solutions \cite{Myung:2006mz,Banerjee:2008gc,Nozari:2008rc,Banerjee:2008du,Banerjee:2009xx,Ma:2017jko,Liang:2017rng,Ghosh:2017odw,Chowdhury:2019odh,Lekbich:2023cjz,Filho:2024zxx,Hamil:2024ppj,Ahmed:2025yuy}. More importantly, these NC corrections to uncharged AdS BHs have been shown to induce interesting features such as phase transitions, \(P\)-\(V\) criticality, and Joule-Thomson expansion \cite{Miao:2015npc,Miao:2016ulg,Graca:2021ker,Wang:2024jlj,Tan:2024jkj,Wang:2025ycl,Sadeghi:2025por}.

The $\kappa$-deformed space-time, $[\hat{x}_i, \hat{x}_0] = ia \hat{x}_i$ and $[\hat{x}_i, \hat{x}_j] = 0$, where $a$ is the $\kappa$-deformation parameter, is a Lie-algebraic type of NC space-time that arises naturally in the framework of Doubly Special Relativity (DSR) \cite{Kowalski-Glikman:2002iba}. The underlying symmetry of DSR is described by the $\kappa$-Poincar\'e algebra \cite{Lukierski:1992dt}, which has also been shown to arise in the low-energy limit of loop quantum gravity (LQG) \cite{Cianfrani:2016ogm}. This $\kappa$-Poincar\'e algebra, in turn, leads to a deformation of the Heisenberg algebra and the standard dispersion relations \cite{Lukierski:1993wx}. Interestingly, the NC versions of the BTZ and Kerr BHs have been shown to be described by a $\kappa$-deformed algebra \cite{Dolan:2006hv,Schupp:2009pt}. Unlike Moyal space-time, the construction of $\kappa$-deformed NC gravity is somewhat intricate due to its underlying Lie-algebraic structure \cite{Rozental:2024eip}. Even so, some studies have been conducted to obtain the $\kappa$-deformations of BH solutions, and their thermodynamics have been investigated. For example, in \cite{Gupta:2013ata,Harikumar:2016bbq}, the $\kappa$-deformed BH solutions have been constructed by employing a realisation approach, which maps NC functions in terms of commutative variables, their conjugates, and the $\kappa$-deformation parameter. Certain aspects of BH in $\kappa$-deformed space-time has been investigated in \cite{Gupta:2013ata,Gupta:2015uga,Juric:2016zey,Harikumar:2016bbq,Gupta:2022oel}, by obtaining the $\kappa$-deformed corrections to Hawking radiation and Bekenstein-Hawking entropy. However, a comprehensive study of the BH thermodynamics in $\kappa$-deformed space-time, along with a detailed analysis of phase transition and criticality, as well as a systematic derivation of the modified first law in extended phase space, has yet to be investigated.

In this work, we study the thermodynamics of the $\kappa$-deformed Schwarzschild-AdS BH in the extended phase space. The NC geometry\cite{Nicolini:2005vd,Kobakhidze:2007jn} and minimal length scale corrections from T-string duality \cite{Nicolini:2019irw}, are shown to induce quantum corrections to the energy-momentum tensor, which in turn induces NC corrections to the standard BH solutions, through Einstein's field equations. Following this general prescription, we first obtain the $\kappa$-deformed energy-momentum tensor for a spherically symmetric matter distribution, by solving the Poisson equation with the $\kappa$-deformed Newton's potential \cite{Rajagopal:2025qyp}. Further we use this $\kappa$-deformed energy-momentum tensor in the Einstein's equation and derive the $\kappa$-deformed corrections to Schwarzschild BH solution. Our analysis shows that the resulting effective $\kappa$-deformed Schwarzschild-AdS solution can serve as a viable phenomenological model for studying $\kappa$-deformations of BH thermodynamics. By interpreting the $\kappa$-deformation parameter $a$ as a new thermodynamic variable in the extended phase space, we formulate the modified first law and the corresponding Smarr relation in $\kappa$-deformed space-time. For the first time, we demonstrate that the NC parameter associated with $\kappa$-deformed space-time can also induce criticality and phase transition behaviour in uncharged Schwarzschild-AdS BHs.

The paper is organised as follows. In Section II, we obtain the metric for the $\kappa$-deformed Schwarzschild-AdS BH. In Section III, we study the thermodynamic properties of this $\kappa$-deformed Schwarzschild-AdS BH by deriving the modified first law and the Smarr relation in Subsection A, and by performing a detailed analysis of criticality and phase transitions in Subsection B. Finally, in Section IV, we summarise our findings and discuss their implications.

\section{$\kappa$-deformed Schwarzschild-AdS black hole}\label{sec1}
 
The NC geometry effectively smears the standard point-like matter distributions and introduces quantum corrections to the energy-momentum tensor in space-time. These NC modifications to the energy-momentum tensor, in turn, deform the standard BH solutions via Einstein's equations \cite{Nicolini:2005vd,Kobakhidze:2007jn}. In this study, we obtain the $\kappa$-deformed corrections to energy density by solving Poisson's equation, using the $\kappa$-deformed Newtonian potential. Employing this $\kappa$-deformed energy density, we then derive the $\kappa$-deformations to the Schwarzschild BH solution within the framework of Einstein's equations.

The $\kappa$-deformed Newtonian potential is obtained as \cite{Rajagopal:2025qyp}
\begin{equation}\label{a1}
  V(r) = -\frac{M}{r}\Big(1-\frac{aM}{12\pi r^2}\Big),
\end{equation}
where the $\kappa$-deformation parameter $a$ is defined in the natural units. Using Eq.(\ref{a1}) in the Poisson's equation $\nabla^2 V = 4\pi {\rho}$, we get the energy density induced by the $\kappa$-deformed NC geometry.
\begin{equation}\label{a1b}
 \rho = \frac{a M^2}{8\pi^2 r^5}
\end{equation}
At the first order in $a$, this energy density decays as $1/r^5$, and vanishes in the asymptotic limit.

The general BH solution of a static, spherically symmetric matter distribution is
\begin{equation}\label{a0}
 ds^2 = -\Big(1-\frac{2m(r)}{r}\Big)dt^2 + \Big(1-\frac{2m(r)}{r}\Big)^{-1}dr^2 +r^2d\Omega^2,
\end{equation}
Satisfying the standard Einstein's equation
\begin{equation}\label{a2a}
 R_{\mu\nu}-\frac{1}{2}g_{\mu\nu}R = 8\pi T_{\mu\nu},
\end{equation}
where $T_{\mu\nu}$ is the $\kappa$-deformed energy-momentum tensor, obeying the conservation law $\nabla^{\mu}T_{\mu\nu}=0$. Since the matter distribution is spherically symmetric, the non-vanishing components of $T_{\mu\nu}$ are defined as $T_{\mu}^{~\nu}=diag(-\rho,p_r,p_t,p_t)$, where $\rho$ is the $\kappa$-deformed energy density, given by Eq.(\ref{a1b}) and $p_r$ and $p_t$ are radial and tangential pressure components, respectively. From the above relations and the conservation equation, we obtain $p_r=-\rho$ and $p_t=-\rho-\frac{r}{2}\frac{d\rho}{dr}$. Substituting Eq.(\ref{a1b}) in these definitions, we get the explicit form of radial and tangential pressures.
\begin{equation}\label{a2c}
\begin{split}
 p_r = -\frac{aM^2}{8\pi^2 r^5}, ~p_t = \frac{3aM^2}{16\pi^2 r^5}.
\end{split}
\end{equation} 
Thus Eq.(\ref{a1b}) and Eq.(\ref{a2c}), constitute the non-vanishing components of $T_{\mu\nu}$. Now using Eq.(\ref{a0}) and the components of energy-momentum tensor in the Einstein's equation Eq.(\ref{a2a}), we obtain $\frac{dm(r)}{dr} = 4\pi r^2 \rho(r)$. Substituting for $\rho$ from Eq.(\ref{a1b}), and solving this, we get the explicit form of the mass function as $m(r)=M\left(1 - \frac{a M}{4\pi r^2}\right)$, and using this in Eq.(\ref{a0}), we obtain the explicit form of $\kappa$-deformed Schwarzschild BH solution, valid up to first order in $a$
\begin{equation}\label{a2}
 ds^2 = -f(r)dt^2 + f(r)^{-1}dr^2 +r^2d\Omega^2,
\end{equation}
where the modified radial function $f(r)$ is
\begin{equation}\label{a1a}
 f(r)=1 - \frac{2M}{r} + \frac{a M^2}{2\pi r^3}.
\end{equation}

\begin{figure}[h!]
\centering
\includegraphics[width=0.475\textwidth]{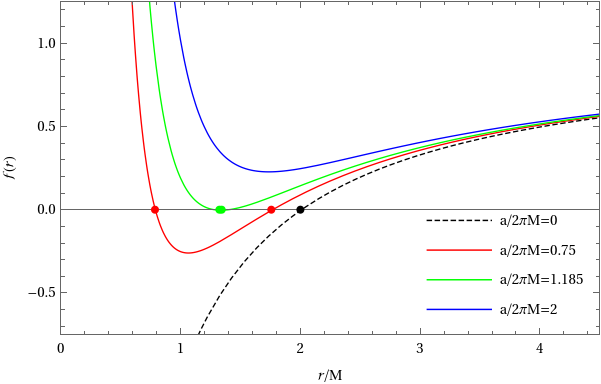}
\caption{Plot of \(f(r)\) against radius for different values of $a$}
\label{fig:fr}
\end{figure}

Fig.(\ref{fig:fr}) shows the behaviour of $f(r)$ against radial distance. The first-order $\kappa$-correction, which depends on $1/r^3$, modifies the Schwarzschild solution. For $a=0$, we recover the standard Schwarzschild BH with an event horizon at $r_+=2M$, as indicated by the dashed line. This first order correction, introduces an outer horizon $r_+$ (event horizon) as well as an inner horizon $r_-$ (Cauchy horzion), as shown by the red curve. Thus the $\kappa$-deformed correction reduces the value of the event horizon compared to the standard Schwarzschild case. As the deformation parameter $a$ increases, the outer horizon shrinks and the inner horizon expands, and at a particular $a$ value, they merge into a degenerate horizon, corresponding to an extremal BH (see green curve). Beyond this, one cannot observe any horizon and results in a naked singularity (see blue curve). A similar behavior due to NC corrections has also been observed for Schwarzschild BH in Moyal space-time \cite{Nicolini:2005vd}.

Interestingly, the $1/r^3$ term also appears in the asymptotic expansion of the regular-Bardeen BH solution, as $f(r) = 1 - \frac{2Mr^2}{(\beta^2 + r^2)^{3/2}} \xrightarrow{~~r>>\beta~~} 1 - \frac{2M}{r} + \frac{6M\beta^2}{2r^3}$ \cite{Ayon-Beato:1998hmi}. Unlike the regular Bardeen BH, the above obtained $\kappa$-deformed BH solution is not regular, as the first order $\kappa$-deformed correction does not remove the singularity of Schwarzschild BH. Similarly the leading order corrections to Schwarzschild BH in Moyal space-time \cite{Lekbich:2023cjz,Filho:2024zxx,Hamil:2024ppj,Ahmed:2025yuy,Miao:2015npc,Miao:2016ulg,Graca:2021ker,Wang:2024jlj,Tan:2024jkj,Wang:2025ycl,Sadeghi:2025por} also, do not remove the singularity. However the Schwarzschild BH in Moyal space-time, incorporating all the higher order NC correction terms are shown to replace this singularity with regular de-Sitter core \cite{Nicolini:2005vd}. In this regard, one need to incorporate all the $\kappa$-deformed higher order terms and investigate separately, whether the higher order $\kappa$-deformation could remove the singularity or not.

The $\kappa$-deformed correction term generates an anisotropic fluid since $p_r\neq p_t$. We also finds that the radial pressure is negative and the tangential pressure is positive and larger than the energy density. From Eq.(\ref{a2c}), we analyse the energy conditions of this deformed BH solution. The effective energy density is positive ($\rho>0$) as long as $a>0$. Similarly, $\rho+p_t>0$ along tangential direction and $\rho+p_r= 0$ is saturated along the radial direction for $a>0$. We also find that $\rho+p_r+2p_t>0$ for $a>0$. Hence the above obtained $\kappa$-deformed Schwarzschild BH solution satisfy the null energy conditions (NEC): $\rho+p_r\geq 0,\rho+p_t\geq 0$, weak energy conditions (WEC): $\rho\geq 0,~\rho+p_r\geq 0,\rho+p_t\geq 0$ and the strong energy conditions (SEC): $\rho+p_r+2p_t\geq 0, \rho+p_r\geq 0, \rho+p_t\geq 0$, respectively. However, it violates the dominant energy conditions (DEC): $\rho\geq 0, \rho\geq|p_r|, \rho\geq|p_t|$, as the tangential pressure component is greater than the energy density. The violation of DEC is also shown in the frameworks of LQG, where the fluid anisotropy and the DEC violation is induced by the $1/r^4$ dependent correction term \cite{Muniz:2024wiv}. Since the $\kappa$-Poincare algebra is shown to arise in the low energy limit of LQG \cite{Cianfrani:2016ogm}, the observed violation of the DEC and the presence of anisotropy in our solution are fully consistent with that of LQG-corrected Schwarzschild BH \cite{Muniz:2024wiv}, thereby showing that our solution naturally fits within the well-motivated quantum gravity framework.

The effective $\kappa$-deformed Schwarzschild--AdS metric can be obtained by including the cosmological constant $\Lambda$ in Eq.(\ref{a2}) and the corresponding $f(r)$ takes the form
\begin{equation}\label{a3}
 f(r) = 1 - \frac{2M}{r} + \frac{a M^2}{2\pi r^3} - \frac{\Lambda r^2}{3}.
\end{equation}
The first order deformation parameter associated with the Moyal space-time \cite{Wang:2024jlj} and the LQG \cite{Wang:2024jtp} are shown to exhibit phase transition and criticality in uncharged Schwarzschild-AdS BHs. This suggests the intriguing possibility that $\kappa$-deformation could also induce phase transition and criticality in uncharged Schwarzschild-AdS BHs, which we explicitly demonstrate in the next section.

\section{$\kappa$-deformed black hole thermodynamics}\label{sec2}

\subsection{First law and Smarr relation}\label{subsec2a}

We study the thermodynamics of the black hole at the event horizon \(r=r_+\). In the extended phase-space formulation of AdS BH thermodynamics, the pressure is defined in terms of the cosmological constant as \cite{Kubiznak:2012wp}
\begin{equation}
P=-\frac{\Lambda}{8\pi}.
\end{equation}
The black hole mass, which is identified with the enthalpy in this framework, can be obtained from the horizon condition \(f(r_+)=0\), namely
\begin{equation}\label{b0}
  f(r_+) = 1 - \frac{2M}{r_+} + \frac{a M^2}{2\pi r_+^3} + \frac{8\pi P r_+^2}{3}=0.
\end{equation}
Solving Eq.(\ref{b0}) up to first order in \(a\), we obtain the mass of the \(\kappa\)-deformed Schwarzschild--AdS black hole as
\begin{equation}\label{b1}
 M = \frac{r_+}{2}\bigg( 1 + \frac{8\pi Pr_+^2}{3} \bigg) + \frac{a}{16\pi} \bigg( 1 + \frac{8\pi Pr_+^2}{3} \bigg)^2.
\end{equation}
The Hawking temperature, obtained from the surface gravity through \(T=\frac{f'(r_+)}{4\pi}\), is given by
\begin{equation}\label{b2}
 T = \frac{1}{4\pi r_+} + 2Pr_+ - \frac{a}{16\pi^2r_+^2} \bigg( 1 + \frac{8\pi Pr_+^2}{3} \bigg)^2.
\end{equation}
For physically relevant BH solutions, the conditions $M>0, r_+>0$ and $T>0$  should be strictly satisfied. Similarly the $\kappa$-deformation parameter in this study represents the length scale associated with the $\kappa$-deformed space-time and hence it should satisfy the condition $a>0$. By imposing these conditions in Eq.(\ref{b1}) and Eq.(\ref{b2}), we obtain constrains on these parameters as $r_+\geq\frac{a}{4\pi},M\geq\frac{3a}{16\pi},P>0$.

The entropy is assumed to follow the Bekenstein--Hawking area law,
\begin{equation}\label{b3}
 S = \frac{A}{4} = \pi r_+^2,
\end{equation}
where \(A\) denotes the horizon area. For a consistent thermodynamic description, these quantities are expected to satisfy the standard first law,
\[
dM = TdS + VdP.
\]
However, using Eqs.(\ref{b1})-(\ref{b3}), we find that
\[
T \neq \left(\frac{\partial M}{\partial S}\right)_P,
\]
which indicates that the conventional first law does not provide a consistent thermodynamic description for the \(\kappa\)-deformed BH.

The apparent inconsistency between the standard first law and the Bekenstein-Hawking area law can be resolved by introducing a modified first law, in which the modification factor depends on the explicit mass dependence of the energy-momentum tensor \cite{Ma:2014qma}. To obtain a consistent thermodynamic description for the \(\kappa\)-deformed BH, we follow the procedure outlined in \cite{Ma:2014qma}, where the modified first law is determined by the explicit form of \(T^0_0(r,M)\). In addition, we treat the NC parameter \(a\) as an independent thermodynamic variable, with \(\Phi_a\) as its conjugate potential. The modified first law then takes the form
\begin{equation}\label{b4}
  W\,dM = TdS + VdP + \Phi_a\,da,
\end{equation}
where
\begin{equation}\label{b5}
 W = 1 + 4\pi\int_{r_+}^{\infty} r^2\frac{\partial T^0_0}{\partial M}\,dr.
\end{equation}
The component \(T^0_0\) of the energy-momentum tensor satisfies the Einstein equation
\begin{equation}\label{a4}
 R^{\mu}_{\nu}-\frac{1}{2}\delta^{\mu}_{\nu}R+\delta^{\mu}_{\nu}\Lambda = 8\pi T^{\mu}_{\nu}.
\end{equation}
From this, one obtains
\begin{equation}\label{a4a}
 \frac{\partial(rf)}{\partial r}-1+r^2\Lambda = 8\pi r^2 T^{0}_{0}.
\end{equation}
Differentiating Eq.(\ref{a4a}) with respect to \(M\) and integrating the resulting expression from \(r_+\) to \(\infty\), we obtain~\cite{Wang:2024jtp}
\[
\int_{r_+}^{\infty} 8\pi r^2\frac{\partial T^0_0}{\partial M}\,dr
=
\int_{r_+}^{\infty}\frac{\partial}{\partial r}\left(r\frac{\partial f}{\partial M}\right)dr \, .
\]
Substituting the explicit form of \(f(r)\) from Eq.(\ref{a3}) into this relation and then using Eq.(\ref{b5}), we obtain
\begin{equation}\label{b6}
 W= 1-\frac{aM}{2\pi r_+^2}.
\end{equation}
In the commutative limit, \(W\to 1\), and the modified first law reduces to the standard form. From Eq.(\ref{b4}), the temperature is then defined as
\begin{equation}\label{b7}
 T = W\bigg(\frac{\partial M}{\partial S}\bigg)_{P,a}.
\end{equation}
Substituting Eqs.(\ref{b1}), (\ref{b3}), and (\ref{b6}) into Eq.(\ref{b7}), one recovers the temperature given in Eq.(\ref{b2}). Similarly, the thermodynamic volume is obtained from the modified first law as
\begin{equation}\label{b8}
 V = W\bigg(\frac{\partial M}{\partial P}\bigg)_{S,a}.
\end{equation}
Using Eqs.(\ref{b1}) and (\ref{b6}) in Eq.(\ref{b8}), we find
\begin{equation}\label{b9}
 V = \frac{4\pi r_+^3}{3}.
\end{equation}
Similarly, Eq.(\ref{b4}) defines the conjugate potential associated with the \(\kappa\)-deformation parameter \(a\) as
\begin{equation}\label{b10}
 \Phi_a = W\bigg(\frac{\partial M}{\partial a}\bigg)_{S,P}.
\end{equation}
Using Eqs.(\ref{b1}) and (\ref{b6}) in Eq.(\ref{b10}), we obtain
\begin{equation}\label{b11}
 \Phi_a = \frac{M^2}{4\pi r_+^2}.
\end{equation}
The Smarr relation for the BH can be derived through a scaling argument based on Euler's theorem. For this purpose, we treat the BH mass as a function of the extended thermodynamic variables \((S,P,a)\), namely \(M=M(S,P,a)\). Under the scaling transformation \(r_+ \to \lambda r_+\), the mass satisfies
\begin{equation}\label{b12}
 \lambda^w M = M(\lambda^x S, \lambda^y P, \lambda^z a),
\end{equation}
where \(w\), \(x\), \(y\), and \(z\) denote the scaling dimensions of \(M\), \(S\), \(P\), and \(a\), respectively. Differentiating Eq.(\ref{b12}) with respect to \(\lambda\), we obtain
\begin{equation}\label{b13}
 w\lambda^{w-1} M = x\lambda^{x-1}S\frac{\partial M}{\partial S} + y\lambda^{y-1}P\frac{\partial M}{\partial P} + z\lambda^{z-1}a\frac{\partial M}{\partial a}.
\end{equation}
From Eqs.(\ref{b0}) and (\ref{b3}), we identify the scaling dimensions as \(w=1\), \(x=2\), \(y=-2\), and \(z=1\). Setting \(\lambda=1\), Eq.(\ref{b13}) reduces to
\begin{equation}\label{b14}
  M = 2S\frac{\partial M}{\partial S} - 2P\frac{\partial M}{\partial P} + a\frac{\partial M}{\partial a}.
\end{equation}
Using the definitions of \(T\), \(V\), and \(\Phi_a\) from Eqs.(\ref{b7}), (\ref{b8}), and (\ref{b10}), we arrive at the Smarr relation for the \(\kappa\)-deformed BH,
\begin{equation}\label{b15}
 WM = 2(TS - PV) + \Phi_a a.
\end{equation}
Although the modified first law given in Eq.(\ref{b4}) is consistent with the Smarr relation above, the quantity \(W\,dM\) is not an exact differential. This feature may have important consequences for defining the differentials of the corresponding thermodynamic potentials.

\subsection{Critical points and phase transition}\label{subsec2b}

From the above expressions, the equation of state of the \(\kappa\)-deformed Schwarzschild-AdS BH can be written as
\begin{equation}\label{c1}
 T = \frac{1}{2\pi v} + Pv - \frac{a}{4\pi^2 v^2} \bigg( 1 + \frac{2\pi Pv^2}{3} \bigg)^2,
\end{equation}
where \(v=2r_+\) denotes the specific volume of the BH \cite{Kubiznak:2012wp}. The critical behaviour of this \(\kappa\)-deformed Schwarzschild-AdS BH can be analyzed using the conditions
\[
\bigg(\frac{\partial T}{\partial v}\bigg)_{a,P}=0,
\qquad
\bigg(\frac{\partial^2 T}{\partial v^2}\bigg)_{a,P}=0,
\]
which determine the inflection point of the equation of state. Solving these equations, we obtain the critical points
\begin{equation}\label{c2}
\begin{split}
 v_c & =\frac{2a(2^{1/3}+1)}{3\pi},\\
 P_c & =\frac{9\pi(2^{1/3}-1)}{8a^2},\\
 T_c & =\frac{3(2^{-1/3}+2^{1/3}-1)}{2a(1+2^{1/3})^2}.
\end{split}
\end{equation}
Unlike the standard charged AdS BH, where the electric charge is responsible for the emergence of critical behaviour, here the \(\kappa\)-deformation parameter \(a\) itself induces criticality in the uncharged Schwarzschild-AdS BH. From Eq.(\ref{c2}), we observe that for a fixed value of \(a\), there exists a unique critical point, and that all critical quantities remain finite only in the presence of non-commutativity.

\begin{figure*}[!]
    \centering
    \begin{subfigure}{0.45\textwidth}
        \centering
        \includegraphics[width=\linewidth]{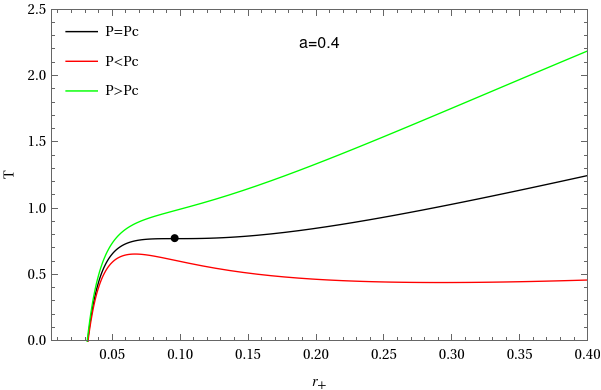}
        \caption{}
        \label{fig:plot1}
    \end{subfigure}
    \hfill
    \begin{subfigure}{0.45\textwidth}
        \centering
        \includegraphics[width=\linewidth]{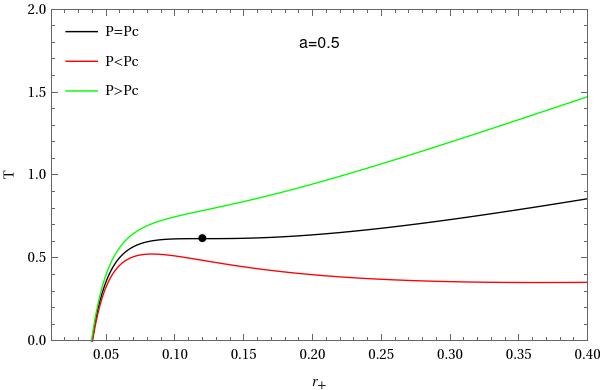}
        \caption{}
        \label{fig:plot2}
    \end{subfigure}
    
    \begin{subfigure}{0.45\textwidth}
        \centering
        \includegraphics[width=\linewidth]{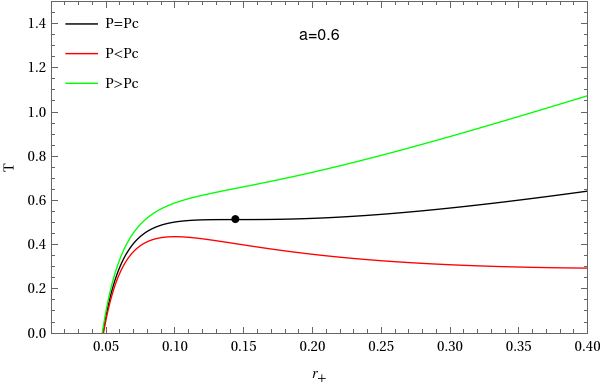}
        \caption{}
        \label{fig:plot3}
    \end{subfigure}
    \hfill
    \begin{subfigure}{0.45\textwidth}
        \centering
        \includegraphics[width=\linewidth]{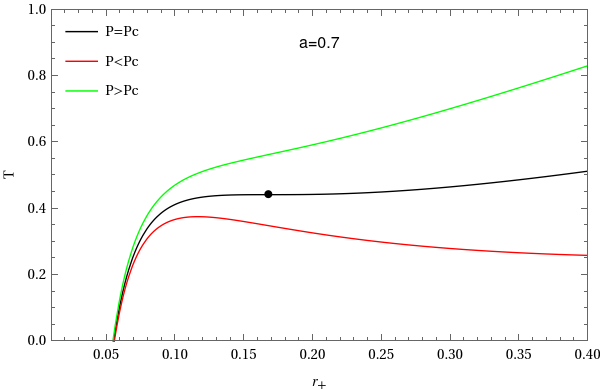}
        \caption{}
        \label{fig:pr}
    \end{subfigure}
 
    \caption{Plot of \(T\) against \(r_+\) for the \(\kappa\)-deformed Schwarzschild-AdS BH for different values of \(a\). Here \(P=0.25P_c\) (red), \(P=P_c\) (black), and \(P=1.5P_c\) (green).}
\label{fig:tp}
\end{figure*}

Fig.(\ref{fig:tp}) shows the isobaric curves of the \(\kappa\)-deformed Schwarzschild-AdS BH for different values of \(a\). For \(P<P_c\), the curves exhibit both a local minimum and a local maximum. In this regime, the system undergoes a phase transition between small and large BHs, analogous to the liquid-gas phase transition in a Van der Waals fluid. As the pressure increases and approaches \(P_c\), these two extrema merge into a single inflection point, represented by the black point in the plot. This marks the critical point, where small and large BH phases coexist. For \(P>P_c\), the temperature varies monotonically with volume along the isobars, and no local extrema appear. Consequently, no phase transition is observed in this region. We also note from Eq.(\ref{c2}) that the critical volume increases linearly with \(a\), whereas the critical temperature decreases inversely with \(a\). This clearly demonstrates how the \(\kappa\)-deformation modifies the critical properties of the Schwarzschild-AdS BH.

As emphasized above, the critical behaviour and the associated phase transition arise entirely due to \(\kappa\)-deformed space-time non-commutativity. From Eq.(\ref{c2}), we obtain the critical ratio
\begin{equation}\label{c3}
 \frac{P_c v_c}{T_c} = 0.370.
\end{equation}
This value is very close to the Van der Waals ratio \(0.375\)~\cite{Kubiznak:2012wp}. Interestingly, the critical ratio is independent of the \(\kappa\)-deformation parameter \(a\), even though the existence of the critical point itself is entirely induced by non-commutativity.

The local thermodynamic stability of the BH can be analyzed in detail through its heat capacity at constant pressure. In this work, we define the isobaric heat capacity as
\[
C_p = T\bigg(\frac{\partial S}{\partial T}\bigg)_{P,a}.
\]
Using Eqs.(\ref{c1}) and (\ref{b3}), we obtain
\begin{widetext}
\begin{equation}\label{c4}
 C_p = \frac{ \pi r_+^2 \big (1728 P \pi^2 r_+^3 + 216 \pi r_+ - 54 a  - 288aP\pi r_+^2  - 384 a P^2 \pi^2 r_+^4 \big)}{864 P \pi^2 r_+^3 - 108 \pi r_+ + 54 a  - 384 a P^2 \pi^2 r_+^4}.
\end{equation}
\end{widetext}

\begin{figure*}[!]
    \centering
    \begin{subfigure}{0.45\textwidth}
        \centering
        \includegraphics[width=\linewidth]{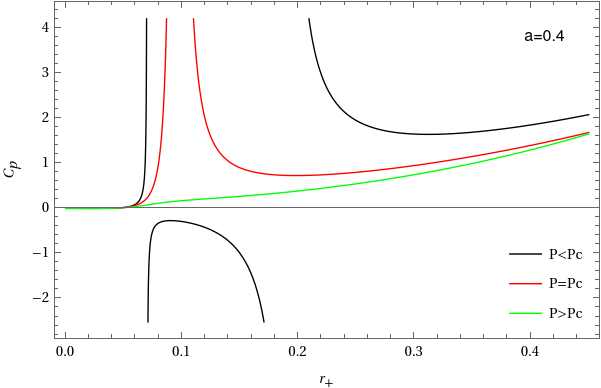}
        \caption{}
        \label{fig:plot1}
    \end{subfigure}
    \hfill
    \begin{subfigure}{0.45\textwidth}
        \centering
        \includegraphics[width=\linewidth]{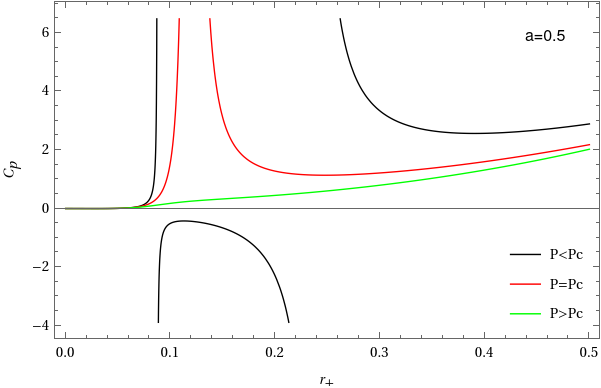}
        \caption{}
        \label{fig:plot2}
    \end{subfigure}
    \begin{subfigure}{0.45\textwidth}
        \centering
        \includegraphics[width=\linewidth]{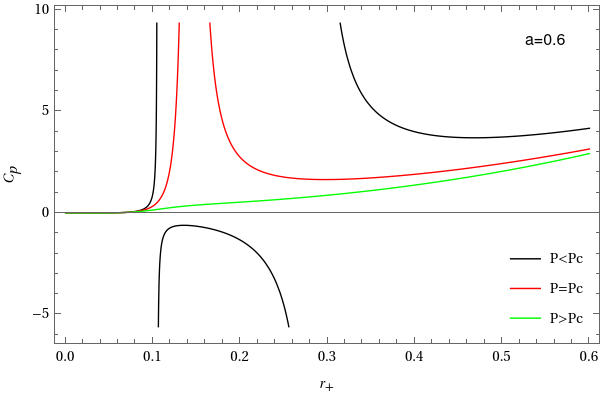}
        \caption{}
        \label{fig:plot3}
    \end{subfigure}
    \hfill
    \begin{subfigure}{0.45\textwidth}
        \centering
        \includegraphics[width=\linewidth]{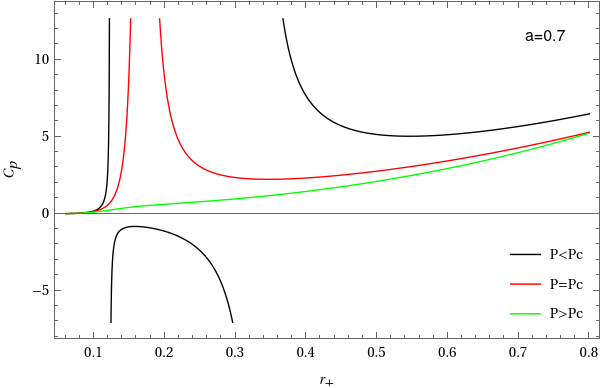}
        \caption{}
        \label{fig:plot4}
    \end{subfigure} 
    \caption{Plot of \(C_p\) against \(r_+\) for the \(\kappa\)-deformed Schwarzschild-AdS BH.}
    \label{fig:cp}
\end{figure*}

Fig.(\ref{fig:cp}) shows the behaviour of the isobaric heat capacity as a function of the horizon radius for different values of \(a\). In the region \(P>P_c\), the heat capacity remains finite over the entire range of \(r_+\), and the corresponding curve is continuous. In this case, one can obtain a stable BH branch characterized by \(C_p>0\). As the pressure is lowered to \(P=P_c\), the heat capacity diverges, signaling the critical point. For \(P<P_c\), \(C_p\) exhibits two divergences, as indicated by the black curves. For small \(r_+\), we first obtain a stable BH branch with \(C_p>0\), which may be identified as a stable small BH. As \(r_+\) increases, \(C_p\) grows and eventually diverges. Beyond this point, \(C_p\) becomes negative, indicating an unstable intermediate BH region. Upon further increasing \(r_+\), \(C_p\) becomes positive again, giving rise to another stable branch with a larger horizon radius, which may be identified as a stable large BH. Thus, in the region \(P<P_c\), the heat capacity analysis clearly supports the existence of a small-to-large BH phase transition discussed above.

\begin{figure}[h!]
\centering
\includegraphics[width=0.5\textwidth]{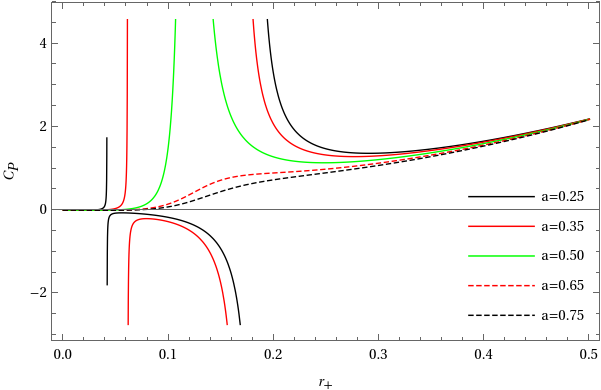}
\caption{Plot of \(C_p\) against \(r_+\) for the \(\kappa\)-deformed Schwarzschild-AdS BH at \(P=P_c\) corresponding to \(a=0.5\), shown for different values of \(a\).}
\label{fig:cpa}
\end{figure}

Fig.(\ref{fig:cpa}) depicts the variation of \(C_p\) with the event horizon radius for different values of \(a\), while keeping the pressure fixed at \(P=P_c\) corresponding to \(a=0.5\). At \(a=0.5\), the heat capacity diverges exactly at the critical point. For \(a<0.5\), the critical radius \(r_c\) decreases in accordance with Eq.(\ref{c2}), and the system exhibits a small-to-large black hole phase transition. As the deformation parameter \(a\) increases, the separation between the two phase transition points gradually decreases. The behaviour changes once \(a\) crosses the value \(0.5\). For \(a>0.5\), the phase transition disappears, as indicated by the dashed curves. Therefore, for \(a>0.5\), only a single stable black hole branch is present, whereas for \(a<0.5\), the system admits a small-to-large black hole phase transition. This behaviour is consistent with that shown in Fig.(\ref{fig:cp}).

\section{Conclusion}


In this work, we have constructed an effective Schwarzschild BH solution within the framework of \(\kappa\)-deformed NC geometry. By using the $\kappa$-deformed Newtonian potential in the Poisson equation, we obtain the $\kappa$-induced energy-momentum tensor, which subsequently, serve as the source term in Einstein's field equations, and generates $\kappa$-deformed corrections to Schwarzschild BH solution. Interestingly, the $1/r^3$ dependent $\kappa$-deformed correction term modifies the horizon structure significantly, by inducing both outer (event) and inner (Cauchy) horizons, respectively. We also observe an extremal BH at the degenerate horizon and beyond this, the horizon cease to exist, leading to a naked singularity. This behaviour has also been observed in NC corrected Schwarzschild BH in Moyal space-time \cite{Nicolini:2005vd}. In general, we observe that the horizon structure of NC-corrected uncharged BHs is remarkably similar to that of standard charged BHs.

The $\kappa$ deformation is shown to induce an effective anisotropic stress-energy tensor, satisfying the null, weak and strong energy conditions. However, it violates the dominant energy condition as the tangential pressure exceeds the energy density. The violation of dominant energy condition and the presence of fluid anisotropy are also observed in LQG corrected BH \cite{Muniz:2024wiv}. Since the $\kappa$-Poincar\'e algebra emerges as the low-energy limit of LQG \cite{Cianfrani:2016ogm}, the DEC violation and anisotropy found in our solution are fully consistent with those of LQG corrected Schwarzschild BHs, thereby placing our effective BH solution within a well-motivated quantum gravity framework.


The $\kappa$-deformed Schwarzschild-AdS BH provides a phenomenological framework for investigating BH thermodynamics in \(\kappa\)-deformed space-time. Using this effective geometry, we carried out a detailed analysis of the thermodynamic properties of the \(\kappa\)-deformed Schwarzschild-AdS BH within the extended phase-space framework. Our results show that \(\kappa\)-deformed space-time non-commutativity significantly modifies the thermodynamic structure of the Schwarzschild-AdS BH and gives rise to new features, such as criticality and phase transitions, which are absent in the commutative limit.

At first sight, the obtained \(\kappa\)-deformed BH solution appears to violate the standard first law due to the leading-order \(\kappa\)-deformation correction. However, by employing the formulation based on the general structure of the energy-momentum tensor \cite{Ma:2014qma}, and by treating the NC parameter \(a\) as an independent thermodynamic variable with conjugate potential \(\Phi_a\), we derived a modified first law containing a correction factor \(W\) that depends explicitly on the \(\kappa\)-deformation parameter. We also obtained a corresponding modified Smarr relation through a scaling argument, consistent with the corrected first law, while preserving the standard Bekenstein-Hawking entropy formula. Similar modifications of the first law and Smarr relation have also been reported in studies of BH thermodynamics in Moyal space-time \cite{Wang:2024jlj,Tan:2024jkj,Wang:2025ycl} and in other quantum gravity frameworks such as loop quantum gravity \cite{Wang:2024jtp}.

We further showed that \(\kappa\)-deformed space-time non-commutativity induces critical behaviour and the associated phase transition between small and large BHs, in contrast to the standard uncharged Schwarzschild-AdS BH, where such behaviour is absent. The critical ratio \(P_c v_c/T_c \simeq 0.370\) is found to be close to the Van der Waals value \(0.375\) and remains independent of the \(\kappa\)-deformation parameter, even though the critical pressure, temperature, and specific volume themselves depend explicitly on \(a\). This universality points to a robust underlying thermodynamic structure associated with the \(\kappa\)-deformed NC correction.

The stability and phase structure of the \(\kappa\)-deformed Schwarzschild-AdS BH were analyzed through the equation of state and the isobaric heat capacity. Our results show that a small-to-large BH phase transition occurs below the critical point. The isobaric heat capacity \(C_p\) diverges at the transition points, separating stable branches with \(C_p>0\), corresponding to small and large BHs, from an unstable intermediate branch with \(C_p<0\). Thus, in addition to inducing criticality, \(\kappa\)-deformed non-commutativity plays a central role in determining the existence of thermodynamic phase transitions and controlling the associated stability properties. 

The emergence of similar minimal-length-scale-induced BH phase transitions has also been observed in the context of Moyal space-time \cite{Wang:2024jlj,Tan:2024jkj,Wang:2025ycl} and LQG \cite{Wang:2024jtp}. In particular, uncharged BHs modified by minimal length scale corrections can exhibit criticality and phase transitions analogous to those of charged BHs in the standard commutative setting. These results highlight how features arising from quantum gravity inspired frameworks, such as LQG and space-time non-commutativity, can manifest themselves in the semiclassical thermodynamic behaviour of BHs. A natural next step would be to extend the present analysis using the Euclidean action approach in \(\kappa\)-deformed space-time. Such an extension would provide a more complete understanding of BH thermodynamics in the \(\kappa\)-deformed framework and may also help clarify its holographic interpretation within the AdS/CFT correspondence.

\section*{Acknowledgment}

VR and PW's research is supported by the National Natural Science Foundation of China (Grant No.~12275080), and the Innovative Research Group of Hunan Province (Grant No.~2024JJ1006). ANK's research is supported by the Croatian Science Foundation Project No. IP-2025-02-8625, \emph{Quantum aspects of gravity}.



\bibliography{ref}

\end{document}